\documentclass[11pt]{article}
\usepackage{latexsym}
\usepackage{amsmath}
\usepackage{amsfonts}
\newtheorem{Def}{Definition}
\newtheorem{The}[Def]{Theorem}
\newtheorem{Pro}[Def]{Proposition}
\newtheorem{Lem}[Def]{Lemma}
\newtheorem{Cor}[Def]{Corollary}
\newtheorem{Rem}{Remark}
\newtheorem{Exa}{Example}

\newenvironment{proof}
{\noindent
{\it Proof.}\ }{\ \(\Box\)}
\def\N{\mathbb{N}}
\def\Z{\mathbb{Z}}
\def\Q{\mathbb{Q}}
\def\r#1{{\rm r}_{#1}}
\def\val#1{{\rm val}_{#1}}

\begin{document}
\title{Construction of regular languages and recognizability of polynomials}
\date{August 27, 1999}
\author{Michel Rigo\\
Institut de Math\'{e}matiques, Universit\'{e} de Li\`{e}ge, \\
Grande Traverse 12 (B 37), B-4000 Li\`{e}ge, Belgium.\\
{\tt M.Rigo@ulg.ac.be}}
\maketitle
\begin{abstract}
A generalization of numeration system in which \(\N\) is recognizable 
by finite automata can be obtained by describing a 
lexicographically ordered infinite regular language. 
Here we show that if \(P \in \Q 
[x]\) is a polynomial such that \(P(\N)\subset \N\) then we can 
construct a numeration system in which the set of representations of 
\(P(\N)\) is regular. The main issue in this construction is to setup 
a regular language with a density function equals to 
\(P(n+1)-P(n)\) for \(n\) large enough.
\end{abstract}

\section{Introduction}
Recently, P. Lecomte and I have introduced in \cite{LR} the concept 
of numeration system on a regular language. A {\it numeration system} 
 is a triple \((L,\Sigma,<)\) where \(L\) is an infinite regular 
language over a totally ordered finite alphabet \((\Sigma,<)\).
The lexicographic ordering of \(L\) gives a one-to-one correspondence 
\(\r{S}\) between the set of the natural numbers \(\N\) and the language \(L\).

For each \(n\in\N\), \(\r{S}(n)\) denotes the \((n+1)^{th}\)  word of \(L\)  with respect to the
lexicographic ordering and is called the {\it \(S\)-representation} of \(n\).

For \(w\in L\), we set \(\val{S} (w)=\r{S}^{-1} (w) \) and  
we call it the {\it numerical value} of \(w\).

When one has a simple method to represent integers, some natural 
questions about ``recognizability" arise. By recognizability, one means 
the following. Let \(S\) be a numeration system and \(X\) be a subset 
of \(\N\). Then \(X\) is said to be {\it \(S\)-recognizable} if 
\(\r{S}(X)\) is recognizable by a finite automaton. Therefore we can 
 consider two kinds of questions.

\(\Diamond \) For a given numeration system \(S\), is it possible to 
determine which 
subsets of \(\N\) are \(S\)-recognizable ?

\(\Diamond \) For a given subset \(X\) of \(\N\), is it possible to find a 
numeration system \(S\) in which \(X\) is \(S\)-recognizable ?

To give a partial but very important answer to the first question, it is 
shown in \cite{LR} 
that arithmetic progressions are always recognizable in any 
numeration system. It is also shown that if \(X\) is recognizable for 
some system \(S\) then \(X+k\) is also \(S\)-recognizable. (These two 
results will be useful in some proofs of this paper.) 

In \cite{R}, we were interested in the second question when \(X\) is the set 
\({\cal P}\) of 
primes. It is shown that \(\r{S} ({\cal P})\) is never 
recognizable for any numeration system \(S\). In this paper, we will 
be mainly concerned by the second question when \(X\) is a 
polynomial image of \(\N\).

For classical numeration systems with integer base, it is well-known 
that the set of the perfect squares is not \(k\)-recognizable
for any \(k\in\N\setminus\{0,1\}\) (see \cite{BHMV} for a 
survey about classical numeration systems). However, in \cite{LR} we 
show  quite easily that the numeration system 
\[S=(a^*b^*\cup a^*c^*,\{a,b,c\},a<b<c)\]
is such that the set \(\r{S}(\{n^2:n\in \N\})\) 
is regular. The choice of the language \(a^*b^*\cup a^*c^*\) was given 
by some density considerations: this language has exactly \(2n+1\) 
words of length \(n\). In view of this result, \hbox{J.-P.~Allouche} asked  
the following question. Is it possible to generalize the result about 
the set of the perfect squares to 
the set \(\{n^k:n\in \N\}\), \(k>2\) ? Moreover, if \(P\) is a polynomial 
belonging to \(\N [x]\) (resp. \(\Z [x]\) or \(\Q [x]\)) such that 
\(P(\N) \subset \N\) then can one find a numeration system such that 
\(P(\N)\) is recognizable ? 

In all these cases, we answer affirmatively. For a given polynomial 
\(P\), we give an explicit method to construct a numeration system 
such that \(\r{S}(P(\N))\) is regular. For this purpose, we show how 
to obtain a regular language which contains exactly \(P(n+1)-P(n)\) 
words of length \(n\) for \(n\) large enough. The construction of regular 
languages with specified density is a problem beyond the concern of 
numeration systems.

The fact that the set of  
primes is never recognizable and that the polynomial images of \(\N\) 
are recognizable give another interpretation of a well-known result  
(see \cite[Theorem 21]{HW}): 
{\it no non constant polynomial \(f(n)\) with integral coefficients can 
be prime for all \(n\), or for all sufficiently large \(n\)}.
\section{Recognizability of polynomials}
Our aim will be to construct a numeration system in which 
\(P(\N)\) is recognizable when \(P\in\Q[x]\) and \(P(\N) \subset 
\N\). 

We will proceed in four steps. First of all, we give an explicit  
iterative 
method to obtain regular languages such that the number of words of 
length \(n\) is exactly \(n^k\) (in \cite{Yu} it is said that such 
languages can be easily obtained). The languages which are given here can 
be interpreted as the basic constructors of our method.

In the three other steps, we 
increase gradually the difficulty. First we consider the case \(P \in 
\N[x]\) which is quite simple since we only deal with the operation of 
addition. Next  we consider 
\(P \in \Z[x]\); here the problem of substraction must be resolved.
Finally,  we have the most general case, \(P \in \Q[x]\) and the 
problem of division. In each of these last three steps, we give an 
instructive short example of construction.

\bigskip
\noindent
{\it \Large i) Languages with density \(n^k\)}
\bigskip

\noindent
First we recall some basic definitions and operations on languages.

\begin{Def}{\rm The {\it density 
function} of a language \(L\subseteq \Sigma^*\) is 
\[\rho_{L}: \N \to \N : n \mapsto \# (\Sigma^n \cap L)\]
where \(\#A\) denotes the cardinality of the set \(A\). 
}\end{Def}

\begin{Def}{\rm If \(x\) and \(y\) are two words of \(\Sigma^*\) then 
the {\it shuffle} of \(x\) and \(y\) is the language \(x \amalg y\) defined 
by 
\[ \{x_{1}y_{1}\ldots x_{n}y_{n} : x=x_{1}\cdots x_{n}, 
y=y_{1}\cdots y_{n}, x_{i},y_{i}\in \Sigma^*, 1\le i\le n, n\ge 1\}.\]
If \(L_{1},L_{2} \subseteq \Sigma^* \) then the {\it shuffle} of the 
two languages is the language
\[L_{1} \amalg L_{2} = \{ w \in \Sigma^*  : w \in x \amalg y, {\rm 
for\ some\ }x \in L_{1},y\in L_{2}\}.\]
}\end{Def}

Recall that if \(L_{1},L_{2}\) are regular then \(L_{1} \amalg L_{2}\) 
is also regular (see for instance \cite[Proposition 3.5]{Ei}).

\begin{Def}{\rm Let \(L\subseteq \Sigma^*\). Then \(\Sigma\) is the {\it 
minimal alphabet of} \(L\) if \(\forall \sigma \in \Sigma\), \(\exists 
w \in L : w=u\sigma v,\ u,v\in \Sigma^*\).
}\end{Def}

We want to construct regular languages \(L_{k}\) such that 
\(\rho_{L_{k}}(n)=n^k\).
The first two languages are, for example, \(L_{0}=a^*\) and 
\(L_{1}=a^+ b^*\). 

To construct a language \(L_{2}\), we first need a language \(M_{2}\) 
such that \(\rho_{M_{2}}(n)=n+1\). We can take \(M_{2}=a^* b^*\). Hence 
\(L_{2}=M_{2} \amalg \{c\}\). Indeed if one considers the 
words of length \(n\) belonging to \(L_{2}\), they are obtained from
\(n\) distinct words of length \(n-1\) belonging to \(M_{2}\) 
 and for each of these words, \(c\) can be 
positioned in \(n\) different places. Thus one has exactly \(n^2\) 
words of length \(n\) in \(L_{2}\). As an example, we have below the 
construction of the nine words of length \(3\), 
\[\begin{array}{ccc}
a^* b^*  & & a^* b^* \amalg \{c\} \cr
aa & \to & aac, aca, caa \cr
ab & \to & abc, acb, cab \cr
bb & \to & bbc, bcb, cbb. \cr
\end{array}
\]
Observe that the letter \(c\) does 
not belong to the minimal alphabet of \(M_{2}\).

To construct \(L_{3}\), we 
simply need a language \(M_{3}\) such that \(\rho_{M_{3}}(n)=(n+1)^2\). This 
can be done using the previously defined languages 
\(L_{0},L_{1},L_{2}\), 
each of them written on a different alphabet,
\[M_{3}=\underbrace{(a^* b^* \amalg \{c\})}_{\rho (n)=n^2}\, \cup \, 
\underbrace{d^+ e^*
\cup f^+ g^*}_{\rho(n)=2n}\, \cup\, \underbrace{h^*}_{\rho(n)=1}.\]
Then we have \(L_{3}=M_{3} \amalg \{i\}\). 

This procedure can be repeated and thus for any 
\(k\ge 2\), \(L_{k}\) can be obtained as a union of previously 
constructed languages and one operation of shuffle with a new letter. 

In the following, the notations \(M_k\) and \(L_k\) will refer to 
the previously constructed languages such that \(\rho_{M_k} (n) = (n+1)^{k-1}\) and 
\(\rho_{L_k} (n) = n^k\).

\begin{Rem}{\rm Let \(u_k\) be the size of the minimal alphabet of 
\(L_k\). The construction of \(L_k\) gives 
\[\left\{\begin{array}{l}
u_0=1,\ u_1=2,\ u_2=3,\cr
u_m={\displaystyle\sum_{k=0}^{m-1}} u_k \binom{m-1}{k} +1 ,\ \forall m\ge 3.\cr
\end{array}\right.
\]
By direct inspection, one can check that \(u_3=9\), \(u_4=26\), \(u_5=90<5!\) and for 
\(n=6,\ldots ,10\), \(u_n<n!\). Let \(m\ge 11\). Since 
\(\binom{m-1}{i}<\binom{m-1}{5}\) for \(i\le 4\); one has 
easily, by recurrence on \(m\), the following upper bound 
\[u_m< \sum_{k=0}^{m-1} k!\, \binom{m-1}{k} = e\, \Gamma(m,1) < e\, 
(m-1)!\]
where \(\Gamma(m,1)\) is the incomplete gamma function defined by
\[\Gamma(a,b)=\int_b^{+\infty} t^{a-1} e^{-t} \, dt.\]
}\end{Rem}

\begin{Rem}{\rm In view of an earlier version of this paper, J.~Shallit 
suggested another construction of a language \(K\) such that 
\(\rho_K (n)=n^k\). It uses the following 
result (see \cite[Section 6.5]{Br})
\[n^k=\sum_{t=0}^k t!\, S(k,t)\, \binom{n}{t}\]
where \(S(k,t)\) are the Stirling numbers of the second kind. The 
language over \{a,b\} with all strings of length \(n\) containing 
exactly \(t\) letters \(b\) is regular and has a density 
\(\rho(n)=\binom{n}{t}\). Therefore a union of such languages on distinct 
alphabets gives the language \(K\). 

This construction is perhaps 
simpler than the construction of \(L_k\) but uses a greater alphabet. 
The size of the minimal alphabet is \(\max_{t=0,\ldots,k} t!\, 
S(k,t)\) and a lower bound is given by \(k!\). We won't use it in the following. 
}\end{Rem}
\bigskip
\noindent
{\it \Large ii) Recognizability of polynomials belonging to \(\N[x]\)}
\bigskip

\noindent
The main idea is that we have to find a regular language such that 
the positions of the first words of each length are the 
values taken by the polynomial.

\begin{Pro}\label{PN} Let \(P \in \N [x]\). If \(P(\N)\subset \N\) then there 
exists a numeration system \(S=(L,\Sigma,<)\) such that \(P(\N)\) is 
\(S\)-recognizable.
\end{Pro}
\begin{proof}
Since the translation by a constant doesn't alter the 
recognizablity of a set, as recalled in the introduction
(see \cite{LR} for details), we can assume that \(P(0)=0\). 
We have to construct a regular language \(L\) such that the 
number of words of length \(n\) is exactly \(P(n+1)-P(n)\). Since 
\(P(n+1)-P(n)\) only contains powers of \(n\) with non-negative integral 
coefficients, the construction of \(L\) can be easily achieved 
by union of languages \(L_{k}\) on distinct alphabets (one has a 
small restriction for the language \(L_0\); we explain it in the 
following example to keep this proof simple). 
To conclude 
the proof, the reader must recall that if a language \(L\) is regular 
then the language \({\cal I} (L)\) formed of the smallest words 
of each length for the lexicographic ordering is still regular 
\cite{Sh}. 
One can check that \(\r{S} (P(\N ))={\cal I} (L)\). 
\end{proof}

\begin{Exa}{\rm Let \(P(x)=2\, x^2+3\, x\). Then
\[P(x+1)-P(x)=4\, x+5.
\]
We consider the language \(L\) which is formed by four copies of 
\(L_1\) and five copies of \(L_0\). 

A very important remark is that with five copies of \(L_0\), {\it we 
obtain five words of any positive length but the only one empty word 
\(\varepsilon\)}. 
So to get rid of this problem we add to our language four new words 
of length \(1\) (we thus add four letters to the alphabet). This 
remark applies for all the following constructions: if one uses 
\(n\) copies of \(L_0\) then add \(n-1\) words of length \(1\) and 
treat the case \(n=1\) separately.

One can check that for \(n\neq 1\), the first word 
of length \(n\) is the \([P(n)+1]^{th}\) word of \(L\) and
\[\r{S}(P(\N\setminus\{1\}))={\cal I}(L\setminus \Sigma).\]
Therefore \(\r{S}(P(\N))\) is regular since we only add one word for 
\(\r{S}(P(1))\) to a regular language.
}\end{Exa}

\begin{Cor} Let \(k\in \N\setminus \{0,1\}\). There exist a numeration 
system \(S\) such that the set \(\{x^k:x\in\N\}\) 
is \(S\)-recognizable. \(\Box\)
\end{Cor}

\bigskip
\noindent
{\it \Large iii) Recognizability of polynomials belonging to \(\Z[x]\)}
\bigskip

\noindent
This lemma gets rid of the problem of the coefficients belonging to 
\(\Z\) instead of \(\N\).

\begin{Lem}\label{PolDif} Let \(k\) and \(\alpha\) be two positive 
integers. There exist a regular language \({\cal L}\) such that 
\(\rho_{\cal L} (n)=n^k-\alpha\, n^{k-1}\) for all \(n\ge \alpha\).
\end{Lem}
\begin{proof}
Assume that \(k\ge 2\). Let \(\Sigma_{k}\) be the minimal alphabet of \(M_{k}\). 
Then \(L_{k}=M_{k}\amalg \{\sigma\}\) where \(\sigma \not\in \Sigma_{k}\). 
For \(i=1,\ldots ,n\), \(L_{k}\) has exactly 
\(n^{k-1}\) words of length \(n\) with \(\sigma\) in position \(i\). 
From this observation, one can check that 
\[{\cal L}=L_{k} \setminus \bigcup_{i=0}^{\alpha-1} \Sigma_k^* \, 
\sigma \, \Sigma_k^i\]
have exactly \(n^k-\alpha \, n^{k-1}\) words of length \(n\) for \(n \ge 
\alpha\). Notice that \(\rho_{\cal L} (n)=0\) if \(n<\alpha\). 

If \(k=1\) then we have to remove the \(\alpha\) first words of each 
length from \(L_1\), 
\[{\cal L}=L_1\setminus [ \underbrace{{\cal I}(L_1)}_{\begin{array}{c}first\ 
words\cr of\ each\ length\cr \end{array}} \cup \ \underbrace{{\cal I}(L_1\setminus {\cal 
I}(L_1))}_{\begin{array}{c}second\ 
words\cr of\ each\ length \end{array}} \ \cup \ldots ]
\]
Notice one more time that \(\rho_{\cal L} (n)=0\) if \(n<\alpha\).
\end{proof}

\begin{Pro}\label{PZ} Let \(P \in \Z [x]\). If \(P(\N)\subset \N\) then there 
exists a numeration system \(S=(L,\Sigma,<)\) such that \(P(\N)\) is 
\(S\)-recognizable.
\end{Pro}
\begin{proof}
We proceed as in Proposition \ref{PN} and consider the 
polynomial \(Q(n)=P(n+1)-P(n)\). Observe that since \(P(\N )\subset 
\N\), the coefficient of the dominant power in \(P\) is positive and 
thus the same remark holds for \(Q\). By adding extra terms of the form 
\(x^j-x^j\), if \(\deg (Q)=k\) we can assume that 
\[Q(x)= x^{i_1+1} - a_{i_1}\, x^{i_1} + 
\cdots + x^{i_r+1} - a_{i_r}\, x^{i_r} + \sum_{l=0}^k b_l\, x^l\]
where \(i_1,\ldots ,i_r \in \{0,\ldots ,k-1\}\), 
\(a_{i_1},\ldots , a_{i_r} \in \N\setminus\{0\}\) and \(b_0,\ldots 
,b_k \in \N\). Let \(\alpha = \sup_{j=1,\ldots ,r} a_{i_j}\). Using Lemma 
\ref{PolDif}, for \(j=1,\ldots ,r\) we construct languages \({\cal L}_j\) such that 
for all \(n\ge \alpha\), \(\rho_{{\cal L}_j} (n)=n^{i_j+1}-a_{i_j}\, n^{i_j}\). 
The reader can construct easily a language \(L\) such \(\forall n\ge 
\alpha\), \(\rho_L 
(n)=Q(n)\) by union of languages \({\cal L}_j\) and \(L_l\). 

If we want to consider the smallest word of each length, as in Proposition 
\ref{PN}, then the 
language \(L\) must contain exactly \(P(\alpha)\) words of length 
at most \(\alpha -1\) (in this case, the first word of length 
\(\alpha\) is the \([P(\alpha)+1]^{th}\) word of \(L\) and its 
numerical value is thus \(P(\alpha)\)). 
This can be achieved by adding or removing a 
finite number of words from the regular language \(L\) (this operation 
doesn't alter the regularity of \(L\)). Thus
\[\r{S}(\{P(n):n\ge \alpha\})={\cal I}(L) \cap \Sigma^{\ge \alpha}.\]

To conclude we have to add a finite number of words for the 
representation of \(P(0),\ldots ,P(\alpha -1)\) and
\[
\r{S}(P(\N))=({\cal I}(L) \cap \Sigma^{\ge \alpha}) \cup \{ 
\r{S}(P(0)), \ldots ,\r{S}(P(\alpha-1)) \}.
\]
\end{proof}
    
\begin{Exa}{\rm Let \(P(x)=x^4-3\, x^2-2\, x+5\). Then
\begin{eqnarray*}
Q(n)=P(n+1)-P(n)&=&4\, x^3+6\, x^2-2\, x-4 \\
 &=& 4\, x^3+5\, x^2+x^2-3\, x+x-4.
\end{eqnarray*}
With four copies of \(L_3\), five copies of \(L_2\) and using Lemma 
\ref{PolDif}, one can construct a regular language \(L\) such 
that\footnote{Here the expression of \(\rho_L (n)\) is very simple since \(3\) and 
\(4\) only differ by one unit (remark that \(4\, n^3+6\, n^2-2\, n-4 =  4\, n^3+6\, 
n^2-3\, n \Leftrightarrow n=4\) and \(4\, n^3+6\, 
n^2-3\, n = 4\, n^3+5\, n^2 \Leftrightarrow n=3\) or \(0\)).}
\[
\rho_L (n)=\left\{
\begin{array}{ll}
4\, n^3+6\, n^2-2\, n-4  & {\rm if}\ n\ge 4\cr
4\, n^3+5\, n^2 & {\rm otherwise}.
\end{array}\right.\]

We have \(P(4)=205\) and the number of words of length at most \(3\) 
belonging to \(L\) is \(214\) thus we remove \(9\) words of length 
at most \(3\) in \(L\). Therefore, the first word of length \(4\) in 
\(L\) is the representation of \(P(4)\) and 
\begin{equation}\label{repres}
\r{S}(\{P(n):n\ge 4\})={\cal I}(L)\cap \Sigma^{\ge 4}
\end{equation}
is a regular subset of \(L\). 
Since \(\{P(0),\ldots ,P(3)\}\) is equal to \(\{1,5,53\}\), we add the second, 
the \(6^{th}\) and the \(54^{th}\) word of \(L\) to (\ref{repres}) to obtain 
\(\r{S}(P(\N))\).
}\end{Exa}

\begin{Exa}{\rm
We begin another example which show how to obtain a correct 
expression for \(\rho_L (n)\) in a trickier situation. Let \(P(x)=x^5-4\, x^3-2\, x^2+8\), then
\[Q(x)=5\, x^4+9\, x^3+x^3-3\, x^2+x^2-12\, x+x-5.\]
To construct a language \(L\), we use five copies of \(L_4\), nine 
copies of \(L_3\) and apply three times Lemma \ref{PolDif}. Thus
\[
\rho_L (n)=\left\{
\begin{array}{ll}
Q(n) & {\rm if}\ n\ge 12\cr 
5\, n^4+10\, n^3-3\, n^2+n-5 & {\rm if}\ 12>n\ge 5 \cr
5\, n^4+10\, n^3-3\, n^2 & {\rm if}\ 5>n\ge 3 \cr
5\, n^4+9\, n^3& {\rm otherwise}.
\end{array}\right.\]    
}\end{Exa}

\bigskip
\noindent
{\it \Large iv) Recognizability of polynomials belonging to \(\Q[x]\)}
\bigskip

\noindent
Finally, we obtain the theorem of recognizability in the general case.

\begin{The} Let \(P \in \Q [x]\). If \(P(\N)\subset \N\) then there 
exists a numeration system \(S=(L,\Sigma,<)\) such that \(P(\N)\) is 
\(S\)-recognizable.
\end{The}
\begin{proof} Let 
\[P(x)=\frac{a_k}{b_k}\, x^k + \frac{a_{k-1}}{b_{k-1}}\, x^{k-1} +  
\cdots + \frac{a_0}{b_0}\]
with \(b_0,\ldots ,b_k,a_k \in \N\setminus \{0\}\) 
and \(a_0,\ldots ,a_{k-1} \in \Z\). Let \(s\) be the least common 
multiple of \(b_0,\ldots ,b_k\). One has
\[P=\frac{P'}{s}\]
with \(P'\in\Z [x]\). By hypothesis \(P(\N)\subset \N\); thus
\(P'(\N)\subset s\, \N\). As in Proposition 
\ref{PZ}, there exist a constant \(\alpha\) and a language \(L'\) such 
that \(\forall n \ge \alpha\),
\[ \rho_{L'}(n) = P'(n+1)-P'(n)=s[P(n+1)-P(n)].
\]
We modify \(L'\) (by adding or removing a finite number of words) to have 
\[
\sum_{i=0}^{\alpha-1} \rho_{L'} (i) = s\, P(\alpha).
\]
It was proved in \cite{LR} that the arithmetic progression \(s\, \N\) is 
recognizable for any numeration system. Let 
\(S'=(L',\Sigma,<)\) then \(L=\r{S'}(s\, \N)\) is a regular language 
such that
\[\sum_{i=0}^{\alpha-1} \rho_{L} (i) =  P(\alpha) \ {\rm and}\ 
\forall n \ge \alpha,\ \rho_L(n) = P(n+1)-P(n).\]
We conclude as in Proposition \ref{PZ}.
\end{proof}

\begin{Exa}{\rm Let 
\begin{eqnarray*}
P(x)&=& \frac{x^4}{3}-2\, x^3+\frac{37}{6}\, x^2-\frac{17}{2}\, x+4 \\
 &=& \frac{1}{3}\, (x-7) x^2 (x+1) + \frac{17}{2} x (x-1)+4.
\end{eqnarray*}
The reader can check easily that \(P(\N)\subset \N\). 
We have \( s=6 \) and 
\begin{eqnarray*}
P'(n+1)-P'(n)&=& 8\, n^3-24\, n^2+46\, n-24 \\ 
 &=& 7\, n^3+45\, n + n^3-24\, n^2 +n-24.        
\end{eqnarray*}
}\end{Exa}
Using seven copies of \(L_3\), \(45\) copies of \(L_1\) 
and applying Lemma \ref{PolDif} twice, 
we construct a language \(L'\) such that
\[\rho_{L'}(n)=\left\{\begin{array}{ll}
6\, (P(n+1)-P(n)) & {\rm if}\ n\ge 24 \cr
7\, n^3 +45 \, n& {\rm otherwise}. \cr
\end{array}\right.\]
The number of words of length at most \(23\) 
in \(L'\) is \(545652\) and \(6\, P(24)=517776\). 
Thus we remove \(27876\) words from \(L'\cap \Sigma^{\le 23}\). In 
this new language lexicographically ordered, we only take the words at 
position \(6i+1\), \(i\in \N\), to obtain the regular language \(L\). Thus the 
\([P(24)+1]^{th}\) word of \(L\) is the first word of length 
\(24\) belonging to \(L\) and 
\[\r{S}(\{ P(n):n\ge 24\})={\cal I}(L) \cap \Sigma^{\ge 24}.\]
To conclude, we have as usual to add a finite number of words for the 
representation of \(P(0),\ldots, P(23)\).

\begin{Rem}{\rm In \cite{LR}, we have studied the problem of changing 
the ordering of the alphabet and we have exhibit some subset \(X\) of \(\N\) and some 
numeration systems \(S\) and \(S'\) which only differ by the ordering 
of the alphabet such that \(\r{S}(X)\) is regular and \(\r{S'}(X)\) 
not. 

This kind of singularity doesn't appear here. 
For a given polynomial \(P\), we have shown how to 
construct a particular numeration system \(S=(L,\Sigma,<)\) such that \(P(\N)\) 
is \(S\)-recognizable. By construction, one can easily check that \(P(\N)\) 
is also \(T\)-recognizable for any system \(T=(L,\Sigma,\prec)\) 
where \(\prec\) is a reordering of \(\Sigma\).
}\end{Rem}

\section{Acknoledgments}
The author would like to thank J.-P.~Allouche and P.~Lecomte for their 
support and fruitful conversations. We also thank J.~Shallit for his 
valuable suggestions.

\end{document}